\documentclass{article}
\newcommand{\be}{\begin{equation}}
\newcommand{\ee}{\end{equation}}
\newcommand{\Su}{S_F^\alpha}
\newcommand{\bw}{\mathbf{w}}
\newcommand{\tf}{\tilde{f}}
\newcommand{\Sk}{S_F^\alpha}
\title{Langevin Equation on Fractal Curves}
\begin{document}
\date{ }
\author{Seema Satin$^{1,2}$ , A.D.Gangal$^3$ \\
\small $1$.Department of Mathematics, University of Pune, Ganeshkhind , Pune-
411007, India .\\
\small $2$. Inter University Center for Astronomy and
 Astrophysics , Ganeshkhind Pune, India. \\
\small $3$. Indian Institute for Science
 Education and Research, (IISER) Pune, India \\
seema.satin@gmail.com, a.gangal@iiserpune.ac.in  }
\maketitle
\begin{abstract}
We analyse a random motion of a particle on a fractal curve, using Langevin
approach. This involves defining a new velocity in terms of mass of the 
fractal curve, as defined in recent work. The geometry of the fractal curve,
hence plays an important role in this analysis. A Langevin equation with a
particular noise model is thus proposed and solved using techniques of the 
newly developed $F^\alpha$-Calculus .
\end{abstract}
\section{Introduction}
Diffusion in disordered media is a topic of immense interest \cite{Bouchaud,
Bouchaud1}.
. There has been a lot of activity in this area recently. Anomalous
diffusion has been studied e.g in \cite{Ott,Soloman,Stapf,Amblard,Hansen}. 
In many of these approaches anomalous diffusion originates due to nonlocalily 
in space or time.

Fractional Langevin equations have been used to describe FBM in several 
references. In \cite{Kou} a theoretical method for subdiffusion based on 
generalized Langevin equation with fractional Gaussian noise is discussed.
In \cite{Jespersen} anomalous transport process is described by a Langevin
equation with Levy noise and corresponding Fokker-Planck equation containing
fractional derivative in space is discussed. In \cite{MetzlerKlafter1} subdiffusion is
established on the basis of an extension of conventional Langevin dynamics to
include long-tailed trapping events. Different methods have been used to solve
 generalized Langevin
equations \cite{porra,lutz}. A Langevin equation for  unbranched 
cracks in 3-D is proposed in \cite{Bouchaud1}.

Here we propose a Langevin equation for particles moving on media, which can
 be modelled as
  fractal curves. 
This equation  is local in space and time, while noise is
 taken to be Levy distributed in general and 
special cases for gaussian and cauchy distributed noise are discussed. This 
Langevin equation on fractal curves involes generalization of velocity for 
motion  of particle on a fractal curve  and 
geometrical concepts for a fractal curve from the recently developed Calculus
on fractal curves \cite{Abhay} . Newly defined fractal integrals and methods of 
solving these integrals is also borrowed from this work. Hence this is a new
formulation for studying nonequilibrium phenonema or anomalous diffusion on
disordered media which can be modelled as fractal curves.  
Solutions of this  Langevin equation in presence of a regular  
Levy stable noise are discussed . We give a method of solving 
this Langevin equation, by using the $F^\alpha$-Calculus developed in 
\cite{Abhay}.
\section{Motion of a particle on a fractal curve} \label{sec:particle}
Let a fractal curve $F$ be embedded in 3-D space, so that, motion of a 
particle on such a fractal curve, can be described by Newton's laws. For 
simplicity we  actually consider the case of a fractal curve embedded in
 $R^2$ (e.g a von Koch curve). 
We consider a particle performing random motion on $F$, then 
such a motion in the continuum limit ( in presence of a noise) can be described by a Langevin equation.

Consider the construction of a von-Koch like curve, carried out  recursively, 
only to a finite stage say $n$. The motion on such a curve, which is made up
 of broken line segments, is then described by Newton's  law. The only
 component of force
$\mathbf{f}$ along these straight  line segments is then relevant. The 
velocity can be quantified  in terms of total distance travelled from the 
initial point along the curve. Since this distance scales locally according to
the $\alpha^{th}$ power of Euclidean distance, a more appropriate quantity to
 define the velocity ( in the limiting case as $n \rightarrow \infty$ ) would
be the change in the values of $S_F^\alpha$, the mass accumulated upto a point
on the curve (see the description below)  with time as the particle moves on
the fractal path. With this motivation we now define '$\alpha$-velocity' of the 
particle as follows:

\begin{eqnarray*}
v^{(\alpha)}(t) &= &\lim_{t \rightarrow t'} \frac{\Su(u(t)) - \Su(u(t'))}{t-t'}  \\
& = & \frac{d}{dt} \Su(u(t)) \\
\end{eqnarray*}
where $\Su(u(t))$ is the rise function as defined in \cite{Abhay},
, which gives
 the mass of the fractal curve $F$, covered upto a certain point on $F$ in 
time $t$. We use the notation $\theta$ to label a point on the fractal curve
$F$ and $J(\theta) = S_F^\alpha(u)$ for the mass of the fractal accumulated 
upto point
$\theta$ as in earlier chapters. 

A similar construction of velocity was introduced in another context 
\cite{cherbit}. The 
crucial difference between that construction and the $v^{(\alpha)}(t)$
 considered here is that, we consider the derivative of $\Su$, rather than 
just $\alpha^{th}$ power of space variable itself.

We now propose the Langevin equation in the overdamped case as: 
\be \label{eq:Langevin}
\frac{d J(\theta(t))}{dt} = \eta(t)
\ee

where $\eta(t)$ is the noise in the system.
The formal solution of the above equation is given by:
\be \label{eq:sol}
J(\theta(t)) = \int_0^t \eta(t') dt'
\ee
\subsection{Model of noise}
In this section we completely follow the noise and corresponding 
renormalization scheme introduced in \cite{Fogedby3}. Therefore, we give a 
brief summary of their model.

We consider a Levy distributed noise $p(\eta) $ (e.g. as given in 
\cite{Jespersen,Fogedby3}) of the form 
\be
p(\eta) =  \mu \eta_0^\mu  \eta^{-1-\mu}
\ee
where $\eta_0$ is a lower cut off, 
introduced to ensure normalization of the distribution $p(\eta)$.

Its Fourier transform is given by
\begin{eqnarray}  \label{eq:noise}
 p(k) &= \langle e^{(-ik\eta)}\rangle = \int d \eta 
\exp(-ik \eta ) p(\eta) 
 =& \exp(-D\eta_0^\mu|k|^\mu)
\end{eqnarray}
where $D$ is a dimensionless geometric factor, $0 < \mu < 2 $ is the scaling 
index. 

In equation (\ref{eq:Langevin}), $\eta(t)$ is the instantly correlated Levy
 white
noise at a particular instant of time. The microscopic steps  $\eta_i$ with 
distribution $p(\eta_i)$  are discrete.
 The corresponding difference equation for
 Langevin equation can be written in terms of discrete time steps 
$\Delta= t/n$, where $n$ is the number of steps or divisions on the time axis.

The Langevin equation is thus discretized as
\be \label{eq:discrete}
\frac{J(\theta_{n+1}) - J(\theta_n)}{ \Delta} = \eta_n 
\ee
 where $J(\theta_n) = J(\theta(t_n))$ and $\eta_n = \eta(t_n)$.

Now, we choose the expression for noise as a stable Levy process of the form 
given in Fourier space as in  \cite{Fogedby3}
\be
P(k,t) = \exp[-D\eta_0^\mu \Delta^{\mu-1} |k|^\mu t ]
\ee
 To keep the coefficent $D$ fixed and eliminate time step $\Delta$, the cut 
off $\eta_0$ has to be renormalized  by, 
$\eta_0^\mu \Delta^{\mu -1}= 1$. Consider equation (\ref{eq:discrete}), and 
take appropriate as follows 
\be
\frac{\langle J(\theta_{n+1}) - J(\theta_n)\rangle_\theta }{ \Delta} = 
\langle \eta_n\rangle_\eta 
\ee
Explicitly evaluating the moment $\langle \eta_n \rangle_\eta$ by using the 
expression for $p(\eta)$ above gives a finite value for $1<\mu<2$, which is
\be
\frac{\langle J(\theta_{n+1}) - J(\theta_n)\rangle_\theta }{ \Delta} = 
const. \eta_0 
\ee
Hence, we can see that  as $\Delta \rightarrow 0$,$\eta_0 \rightarrow \infty$ 
i.e. the cut-off moves to infinity, which hold for $\eta_0^\mu \Delta^{\mu-1}=1
$ in this range for $\mu$. For $0 < \mu <1$, in order for the 
renormalization $\eta_0^\mu \Delta^{\mu -1}=1$, to hold, 
$\eta_0 \rightarrow 0 $ for $\Delta \rightarrow 0$. This renormalization
is same as mentioned above in \cite{Fogedby3} for ordinary Langevin equation.

Further, we denote $D \eta_0^\mu = D_1$.
\subsection{Solution of the Langevin equation}

We define the function $\delta_F^\alpha$ which is analogous to dirac delta
function with respect to $F^\alpha$-integrals. Thus
\[\delta_F^\alpha(\theta(t) - \theta(t'))\ = \delta(J(\theta(t))- J(\theta(t'))
\]
The formal solution of the Langevin equation (\ref{eq:Langevin}) is given by
 equation
(\ref{eq:sol}). The associated distribution can now be found as follows. 

Let $p(\theta,t)$ denote the probability distribution for a particle located
at  point $\theta$ on
$F$ at time $t$. Thus
\be \label{eq:deltafunction}
p(\theta_0,t) = \langle \delta_F^\alpha (\theta_0 - \theta(t)) \rangle_\theta
\ee

Now the  Fourier Transform as defined in appendix is given by
\be
\tilde{f}(\psi) =  \int_{C(-\infty,\infty)} e^{-i J(\psi)
J(\theta)} {f}(\theta) d_F^\alpha \theta
\ee
and the inverse Fourier Transform as

\be
f(\theta) = \frac{1}{ 2 \pi}\int_{C(-\infty,\infty)} e^{i J(\psi) J(\theta)} 
\tilde{f}(\psi)d_F^\alpha \psi
\ee
 then, $\delta_F^\alpha$ can be defined in terms of $F^\alpha$ integral 
(this can be obtained easily by applying conjugacy ( 
\cite{Abhay} to the $\delta$-function in ordinary case) as:
\be
\delta_F^\alpha(\theta) = \frac{1}{2 \pi} \int_{C(-\infty,\infty)} 
\exp(i J(\theta) J(\psi)) d_{F}^\alpha \psi 
\ee
thus equation (\ref{eq:deltafunction}) becomes
\[
p(\theta,t) = \langle \frac{1}{2 \pi} \int d_{F}^\alpha \psi  
\exp(i J(\psi)[J(\theta) -J(\theta'(t))]) \rangle_{\theta'}
\]
or
\[
p(\theta,t) =  \int d_F^\alpha \theta' p(\theta') (\frac{1}{2 \pi} \int
 d_F^\alpha \psi  \exp(i J(\psi)[J(\theta) -J(\theta'(t))])) 
\]

The integrals can be interchanged and 

\[
p(\theta,t) =  \frac{1}{2 \pi} [  \int d_F^\alpha \psi \int
d_F^\alpha \theta'  p(\theta') \exp(i J(\psi)J(\theta)\exp(-i J(\psi) 
J(\theta'(t))] 
\]
or 
\be
p(\theta,t) =  \frac{1}{2 \pi}  \int d_F^\alpha \psi 
 \exp(i J(\psi)J(\theta) \langle \exp(-i J(\psi) 
J(\theta'(t))\rangle_{\theta'}
\ee
Hence we can see that
\[p(\theta,t) = \frac{1}{2 \pi}\int d_F^\alpha \psi \exp(i J(\psi)
J(\theta)) \tilde{p}(\psi,t)\]
Thus, taking the inverse Fourier transform of the above,
\[\tilde{p}(\psi,t) = \int d_F^\alpha \theta 
\exp(-i J(\psi)J(\theta)  p(\theta,t)\]
or
\[\tilde{p}(\psi,t) =  \langle \exp(- i J(\psi) J(\theta)) \rangle_{\theta} \]
Substituting the value of $J(\theta)$ from equation (\ref{eq:sol})
\be
\tilde{p}(\psi,t) = \langle \exp(- i J(\psi) 
\{ \int_0^t \eta(t') dt'\}) \rangle_{\eta}
\ee
Discretizing the integral in the above equation
\be \label{eq:psi}
\tilde{p}(\psi,t) =   \prod_{n=0}^N  \langle 
\exp[ -i J(\psi) \eta(t_n) \Delta \rangle \mbox{ where } t_n = n \Delta
\ee
such that  the interval $[0,t]$ is divided into $N$ equal parts and 
$\Delta = (t-0)/N$. 

We now assume a model of noise $\eta(t)$ which is given by \ref{eq:noise}. 

Comparing  equation (\ref{eq:psi}) with the noise model (\ref{eq:noise}) above
 we can write
\be
\tilde{p}(\psi,t) =   \prod_{n=0}^N \exp[ - D_1 |J(\psi) \Delta|^\mu]
\ee
Using the renormalization $D_1 \Delta^{\mu-1} \rightarrow D$ and
 reintroducing the integral in the above equation: 
\[
\tilde{p}(\psi,t) =    \exp[ - D |J(\psi)|^\mu\int_0^{t} dt'\}]
\]

\be
\tilde{p}(\psi,t) =   \exp[ - ( D |J(\psi) = \Sk(k)|^\mu t)]
\ee

For $\mu =2$ we get
\be
\tilde{p}(\psi,t) =   \exp[ - (D \Sk(k)^2 t)]
\ee
The Fourier transform for the above equation gives
\be \label{eq:gaussian}
p(\theta,t) = \frac{1}{\sqrt{2 \pi D t}} \exp{- (\frac{\Su(u)^2}{2 Dt})} 
\ee 
We see that the equation (\ref{eq:gaussian}) is the same as that obtained for
 the solution of diffusion equation on fractal curve as discussed in 
\cite{Seema}.
\section{Results}
In this paper we have proposed a Langevin equation for random motion of a
 particle on a fractal curve. The Langevin equation we propose is local in 
space and time, while noise is taken to be Levy distributed.  
Special case for gaussian distributed noise is discussed. This 
Langevin equation on fractal curves involves defining velocities and 
geometrical concepts for a fractal curve from the Calculus
on fractal curves developed in \cite{Abhay} .  
Our consideration demonstrates that the 
framework we have introduced is suitable for  studying nonequilibrium 
phenonema or anomalous diffusion on fractally
disordered media, which can be modelled as fractal curves (such as backbone of
 a percolating cluster, polymer chain etc.)  

For gaussian distributed white noise we have shown that the probability 
distribution of the space variable obtained from the Langevin equation is
the  same
as that obtained by solving a Fokker-Planck equation on a fractal curve, as 
done in \cite{Seema}. Hence the two approaches are equivalent as
 in ordinary space, for this special case.

\section*{Appendix 1}

\textbf{Review of Calculus on Fractal Curves}

For a set F and a subdivision $P_{[a,b]}, a<b$, $[a,b]
\subset [a_0,b_0]$ let $\mathbf{w}:[a,b] \rightarrow F$, then we define the 
mass function as follows:

\be \gamma^{\alpha}(F,a,b) = \lim_{\delta \rightarrow 0} \inf_{P_{[a,b]}:|P|
 \leq \delta} \sum_{i=0}^{n-1} 
 \frac{|\mathbf{w}(t_{i+1}) - 
\mathbf{w}(t_i)|^\alpha}
{\Gamma(\alpha +1)}  \label{eq:sigma}\ee
where $|\cdot|$ denotes the euclidean norm on $\mathbf{R^n}$, $1 \leq \alpha
\leq n$ and $P_{[a,b]} = 
\{a=t_0,\ldots,t_n=b\}$. 

The staircase function, which gives the mass of the curve upto a certain point
on the fractal curve $F$ is defined as
\be
S_F^{\alpha}(u) = \left\{ \begin{array}{ll}
	\gamma^{\alpha}(F,p_0,u) & u \geq p_0 \\
	- \gamma^\alpha(F,u,p_0) & u < p_0
		\end{array}
	\right. 
	\label{eq:staircase_function} \ee
where $u \in [a_0,b_0]$.

A point on the curve $\mathbf{w}(u) \equiv \theta $ and $S_F^\alpha(u) \equiv
J(\theta)$.
The $F^\alpha$ derivative is defined as:

\be (D_F^\alpha f)(\theta)= F \mbox{-}\lim_{\theta' \rightarrow \theta} 
\frac{f(\theta')-f(\theta)}
{J(\theta')-J(\theta)} \label{eq:derivative}\ee
 
The $F^\alpha$-integral is also defined and is denoted by
\be \int_{C(a,b)} f(\theta) d_F^\alpha \theta
\ee 

\section*{Appendix 2}

\textbf{The Fourier Transform}

From the definition of conjugacy (\cite{Abhay})
\be \label{eq:defcon}
\phi[f] (\Su(u)) = f(\mathbf{w}(u))
\ee
The  Fourier Transform on the real line, for a function $g(v)$, is defined
 by
\be
g(v) = \int_{-\infty}^{\infty} \tilde{g}(y) \exp(-ivy) dy 
\ee
and the inverse Fourier Transform is
 
\be \label{eq:invtrans}
\tilde{g}(y) = \frac{1}{2 \pi}\int_{-\infty}^{\infty} g(v) \exp(ivy) dy 
\ee
In the case of a parametrizable fractal curve $F$, which is obtained
by a fractalizing transformation on an interval of the real line, we propose
 that the Fourier space can also be obtained by the same fractalizing 
transformation on the interval of a real line. The interval may be $(-\infty,
\infty)$.

 Let $\tilde{g} = \phi[\tf]$ and $g = \phi[f]$ also $v=\Su(k)$ and $y
= \Su(u)$.

We use the notation $J(\theta) = \Su(u)$ and $J(\psi) = \Su(k)$, where
$\theta = \bw(u)$ and $\psi = \bw(k)$.

Then taking Fourier transform of the LHS of equation (\ref{eq:defcon}) one can
 write
\begin{eqnarray} \label{eq:fourier}
\tilde{\phi[f]} (v=J(\psi)) &=& \int_{-\infty}^\infty \phi[f](y =J(\theta))
\exp(-iyv)dy \nonumber \\
& = &  \int_{C(-\infty,\infty)} f(\theta) \exp(-iJ(\theta) v)d_F^\alpha \theta 
\end{eqnarray}
 where 
\[C_(-\infty,\infty) = \lim_{a \rightarrow -\infty, b \rightarrow \infty} 
C_(a,b) \]
and 
\begin{eqnarray} \label{eq:fourier1}
\phi[\tf] (v=J(\psi)) & = & \phi [ \int_{-\infty}^\infty f(y =J(\theta)) 
\exp(-iyv)dy] \nonumber \\
& = &  \int_{C(-\infty,\infty)}f(\theta) \exp(-iJ(\theta) v)d_F^\alpha \theta
\end{eqnarray}
Comparing equations (\ref{eq:fourier}) and (\ref{eq:fourier1}), we can write
\[\tilde{\phi[f]} = \phi [\tilde{f}] \]
Also one can define the action of $\phi$ in Fourier space as
\be
\phi[\tilde{f}](v = J(\psi)) = \tf(\psi)
\ee
Now using conjugacy one can rewrite equation (\ref{eq:fourier}) as 

\be
\tf(\psi) =  \int_{C(-\infty,\infty)}f(\theta)
\exp(-i J(\theta) J(\psi) ) d_F^\alpha \theta
 \ee
Similarly, inverse transform of the above can be obtained from 
equation (\ref{eq:invtrans}), which can be 
written as

\be
f(\theta) = \frac{1}{2 \pi} \int_{C(-\infty,\infty)}\tf(\psi)
\exp(i J(\theta) J(\psi) ) d_F^\alpha \theta
 \ee

$\bullet$

\end{document}